# Spin dependent transport in ferromagnetic/superconductor/ferromagnetic single electron transistor

Dawei Wang, Jia G. Lu*

*Department of Electrical Engineering and Computer Science & Department of Chemical Engineering and Materials Science, University of California, Irvine, CA 92697*

Ferromagnetic single electron transistors with Al islands and orthogonal ferromagnetic leads (Co) are fabricated using ebeam lithography followed by shadow evaporation techniques. *I-V* characteristics exhibit typical single electron tunneling effects. Transport measurements performed in external magnetic field show that, when the two ferromagnetic leads are in antiparallel configuration, spin imbalance leads to a suppression of superconductivity.

## INTRODUCTION

For decades, electronic devices have been relying on the transport of electronic charge. The electron spin degree of freedom, however, is often ignored. A new technology called spintronics (or magneto-electronics) has emerged since the discovery of giant magnetoresistance effect in 1988.[1,2] Today, GMR read head has become the key in high-density magnetic information storage. At the same time, much progress has been made in making magnetic random access memory based on tunneling magnetoresistance (TMR) effect.[3,4] The ability to manipulate electron spin states is crucial to extremely high-density information storage, electron-spin-based quantum computing, and magneto-electronic sensors. As the trend of device miniaturization continues, such applications require fundamental understanding of the underlying physics of spin dependent transport in nanoscale structures.

Recent advances in nanofabrication make it possible to achieve double tunnel junction systems with very small junction areas, called single electron transistors. It consists of a small metallic island that is weakly coupled to two bias leads through small capacitance tunnel junctions, and capacitively coupled to a gate electrode. The total capacitance $C_\Sigma$ (the sum of two junction capacitances and the gate capacitance $C_g$) is so small that the charging energy ($E_c \equiv e^2/2C_\Sigma$) required to add a single electron onto the island becomes the dominant energy at low temperatures and low bias voltages. Thus, the charging energy barrier prevents electrons from tunneling on or off the island. This situation is referred as the Coulomb blockade (CB). Several interesting effects have been theoretically predicted for spin-dependent transport in ferromagnet (FM)/metal/FM and FM/superconductor/FM single electron transistors. They include enhanced TMR in the CB region, spin accumulation, and superconducting gap suppression.[5] For a normal metal island with long spin diffusion length, when the magnetizations of two ferromagnetic leads are in antiparallel configuration (see Fig. 1a), a non-equilibrium spin density (spin accumulation) is induced on the central electrode (island) due to the imbalance of the tunneling current formed by spin-up and spin-down electrons. When the island is superconducting, the resulting non-equilibrium spin density strongly suppresses the superconductivity with increasing bias voltage.[5,6]

In this paper, we report our experimental results on a ferromagnetic single electron transistor (FMSET) with Al as the superconductor island and Co as the FM leads. From the transport measurements, we present the observation manifesting the spin accumulation effect leading to the suppression of the superconducting gap.

## EXPERIMENT

Our fabricated FMSET consists of two Co leads coupled to an Al island through $Al_2O_3$ tunnel junctions with small capacitance and large tunnel resistance ($R_T = R_1 + R_2$ is much higher than resistance quantum, *i.e.* $R_T >> h/e^2$). Fig. 1(a) is a schematic of a FMSET and Fig.1(b) shows the Atomic Force Microscope (AFM) image of a fabricated device. The FM leads are fabricated perpendicular to the Al island in order to achieve a stabilized antiparallel configuration.[7-9] A gate electrode patterned 1.6 µm away from the island can be used to tune the potential on the island.

The device is fabricated using ebeam lithography followed by angled deposition techniques. In our experiment, a bi-layer ebeam resist is used. The bottom layer of 480 nm copolymer of methylmethacrylate (MMA) and methacrylic acid (MAA) is first spun coated on a Si chip. After baking at 180º C for 30 minutes, top resist layer polymethylmethacrylate (PMMA) of ~100 nm thick is coated and baked at 180º C for 20 minutes. The key point in the ebeam lithography process is to get a narrow linewidth in the top resist layer and at the same time a large undercut in the bottom resist. The ratio between top linewidth and undercut needs to be at least 1:6 in order to achieve successful shadow evaporation to form the junctions. The exposure is done on ebeam writer (NPGS system) at an accelerating voltage of 12 kV. Afterwards, the sample is developed, and followed by metal deposition in an ebeam evaporator (CHA ebeam evaporator). 25nm Al is first deposited forming the central island. The evaporation angle is set such that the lines caused by this evaporation are offset from the actual Co electrodes. After oxidation, 40nm Co is evaporated towards the island, forming the two tunnel junctions.[10] The difference in the widths of the Co leads (50 nm and 90 nm, respectively) results in a difference of the coercive fields. Magnetic field dependent transport measurement is carried out in a He3 cryostat (Jannis H3-3-SSV/Magnet) at a base temperature of 0.27 K.

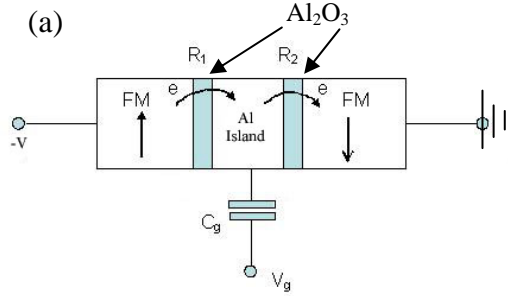

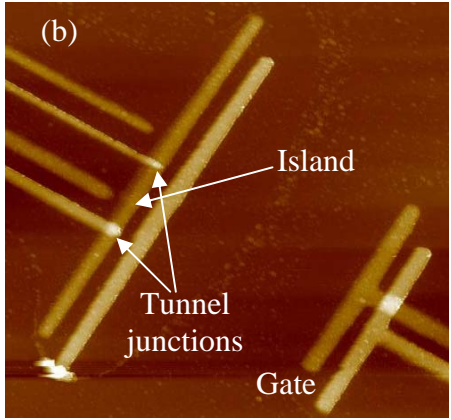

FIG. 1. (a) Schematic of a FMSET with Al island and Co leads. (b) AFM image of a fabricated FMSET. The two Co leads (perpendicular to the Al island) are 50 nm and 90 nm in width, respectively.

## RESULTS AND DISCUSSION

$I$ - $V$ measurements of the device show typical single electron tunneling behavior. At low temperature, the $I$ - $V$ curve shows a gap region at low bias voltages (as shown in Fig. 2). This high resistance gap is due to both Coulomb blockade ($2E_c/e$) and superconducting energy gap ($2\Delta/e$) of the Al island. At 1.95 K, above the critical temperature of the Al film (~1.46 K for this device), the gap due to superconductivity disappears, but the $I$-$V$ is still slightly nonlinear due to the CB effect at $V_g = 0$. From these measurement results, the sample parameters can be estimated, yielding total junction resistance $R_T \approx 4.3$ M$\Omega$, total capacitance $C_\Sigma \approx 0.5$ fF, and a superconducting gap $\Delta \approx 220$ μeV.

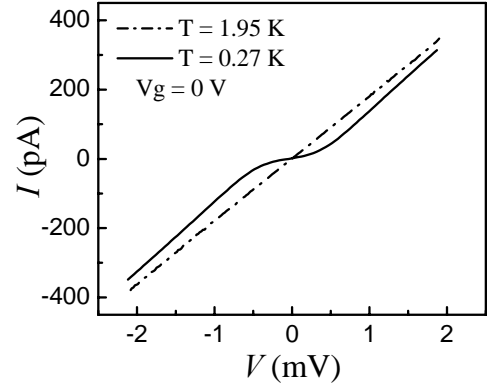

FIG. 2. $I$ - $V$ characteristics of the FMSET. $I$ - $V$ curves obtained at different temperatures. At temperature (0.27 K) below the critical temperature of Al (~1.46 K), a clear gap can be observed; At $T$=1.95 K, the superconducting gap is destroyed.

Measurements performed in magnetic fields applied along the leads exhibit several interesting results. In Fig. 3, as the field is lowered, the current decreases quickly between -5.2 and +5.2 kOe. Beyond this range, the current gradually levels. Such bell shaped $I$ - $H$ curve is a result of the direct influence of the external magnetic field on the superconducting gap. And it yields a critical magnetic field ($H_{C||}$) of ±5.2 kOe for this sample. $H_{C||}$ at different temperatures is extracted from the temperature dependent $I$ - $H$ measurement and plotted in Fig. 4. The experimental result fits well with theory (solid line is the fitting result): $H_{c||}(T) = H_{c||}(0)[1-(T/T_c)^2]/[1-(T/T_c)^4]$, where $H_{c||}(T)$ is the critical field of the thin film as a function of temperature $T$, $H_{c||}(0)$ is the critical field at $T$ =0, and $T_c$ is the critical temperature of the superconductor.[11]

Another important feature of Fig. 3 is that, when the magnetic field sweeps from -6 kOe to +6 kOe, a small current peak appears between 0.6 kOe and 1.4 kOe; correspondingly, when the field reverses from +6 kOe to -6 kOe, a similar peak appears between -0.6 kOe and -1.4 kOe (indicated by the down arrows). We believe that this is caused by the magnetization switching in the Co electrodes from parallel to antiparallel configurations due to the different coercive fields of the two electrodes. Because of the long spin diffusion length of Al,[12] spin flip process can be neglected. The probability of electron tunneling from the source electrode depends on the number of available states of the same spin direction in the drain electrode. When the two Co leads are in antiparallel configuration, the electrons in the source electrode with majority spins, are easier to tunnel onto the island, however encounter a higher resistance to tunnel off the island. This results in a spin imbalance in the island, leading to a shift in the chemical potential for different electron spins. For a normal metal island, the chemical potential difference, $\delta\mu$, is calculated to be proportional to the voltage drop across the device and the spin polarization of the FM electrodes.[5] For the superconducting Al island, this difference gives rise to an imbalance of spin-up and spin-down electrons, which reduces the formation of cooper pairs, and consequently suppresses superconductivity.[6] As shown in Fig. 3, at high fields, the magnetization of the electrodes saturates and aligns parallel to the external field. As the magnitude of the applied field decreases and field changes sign, the wider Co electrode (90 nm) switches first before the thinner electrode (50 nm) switches. It is in this 800 Oe window where the magnetizations are in an antiparallel state, spin accumulation causes a suppression of the superconducting gap, resulting in an increase in current.

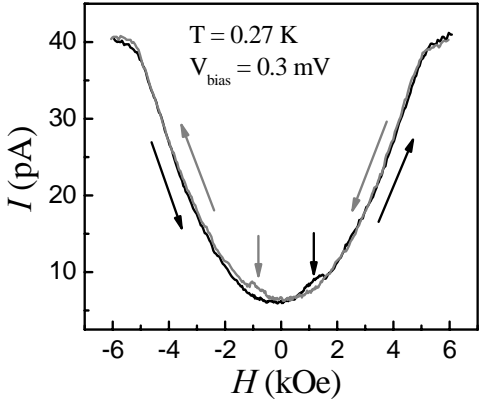

FIG. 3. Current vs. sweeping field. The arrows along the curve indicate the sweeping direction of the applied magnetic field.

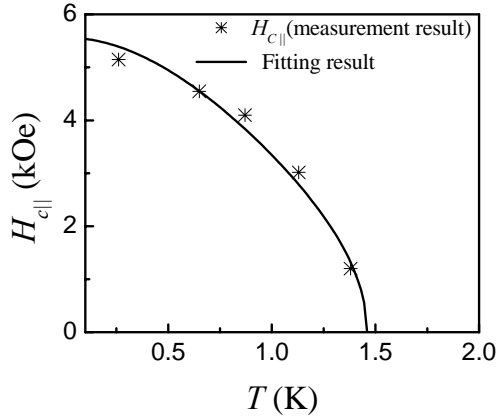

FIG. 4. Critical fields measured at different temperatures. Solid line is the fitting result of $H_{C||}$ vs. $T$.

In conclusion, FMSET is fabricated using Co as leads and Al as the superconducting island. $I$ - $V$ characteristics of the device shows typical single electron tunneling behavior, and basic parameters of the device is extracted from the measurement results. The superconducting gap suppression due to spin accumulation on the Al island is observed when the ferromagnetic leads switch from the parallel to the antiparallel configuration.

## ACKNOWLEDGEMENT

The authors thank Professor Gerd Bergmann for valuable discussion. This project is supported by NSF NIRT grant #DMR-0334231. The device fabrication was done at UC Irvine Integrated Nanosystems Research Facility.